\documentclass[prb,11pt]{revtex4-1}
\usepackage{amssymb}%
\usepackage{amsmath}%
\usepackage{amsfonts}%
\usepackage[dvips]{graphicx,color}
\usepackage{graphics}
\usepackage{subfigure}

\begin{document}
\title{Semiconductor quantum wells with BenDaniel - Duke boundary conditions: approximate analytical results}
\author{Victor Barsan}
\email[Corresponding author: ]{vbarsan@theory.nipne.ro}
\author{Mihaela-Cristina Ciornei}
\affiliation{IFIN-HH, Aleea Reactorului no. 30, Magurele 077125, Romania}
\date{\today}

\begin{abstract}
  The Schr\"{o}dinger equation for a particle moving in a square well potential with BenDaniel - Duke boundary conditions is solved. Using algebraic approximations for trigonometric functions, the transcendental equations of the bound states energy are transformed into tractable, algebraic equations. For the ground state and the first excited state, they are cubic equations; we obtain simple formulas for their physically interesting roots. The case of higher excited states is also analyzed. Our results have direct applications in the physics of type I and type II semiconductor heterostructures.
\end{abstract}

\maketitle

\section{Introduction}

This paper is devoted to the study of a simple problem of quantum mechanics:
the movement of a particle with position-dependent mass (PDM) in a finite square
well. The physical situation modelled by such a potential can be an electron in the conduction band of a semiconductor moving in a heterostructure, let us say $BAB$, i.e. a thin layer of a semiconductor $A$ sandwiched between two somewhat larger semiconductors of identical composition, B (Ref. \cite{Ihn}, p.66). So, simple as it is, this problem involves two important issues: the position-dependent mass (PDM) quantum physics and semiconductor heterostructures. Let us shortly comment on these points.

As the effective mass of a charge carrier in a semiconductor depends on the
charge carrier - lattice interaction, it changes if the lattice composition or
the symmetry changes. So, excepting the case of a charge moving in a perfect crystal, the effective mass of an electron or hole is, rigorously speaking, position-dependent.

The roots of the position dependent effective mass concept are to be found in the
pioneering works of Wannier (1937) and Slater (1949) (see Ref. 1 in Ref.$\ $\cite{vR}).
Recent papers give explicit methods to obtain explicit solutions of the
Schr\"{o}dinger equation with PDM, for various forms of this
dependence and for several classes of potentials \cite{Levai}, \cite{Nikitin}, \cite{SAE}.

However, in practical situations usually encountered in the physics of
semiconductor junctions of two materials, $A$ and $B,$ the simplest and more
popular form of position dependence of the effective mass is a step function:
the effective mass has a constant value in the material $A$ and another,
constant value, in the material $B.$ In such a case, the most convenient
approach for obtaining the wave functions or the envelope functions in a
heterostructure - for instance, a quantum well (QW) or quantum dot (QD)- is to
solve the Schr\"{o}dinger equation with BenDaniel - Duke boundary conditions for the wave function \cite{Bastard}, \cite{Ihn}.

 The transition from the complex problem of a real
semiconductor (for instance Kane theory), to the simple problem of a particle moving in a square well
with BenDaniel - Duke boundary conditions is indicated, for instance, in Chapter III of
Bastard's book \cite{Bastard}.  This simple problem provides however a realistic description of states near the high-symmetry points in the Brillouin zone of a large class of semiconductors. "It [i.e. "the simple problem"] often leads to
analytical results and leaves the user with the feeling that he can trace
back, in a relatively transparent way, the physical origin of the numerical
results." (Ref.$\ $\cite{Bastard}, p. 63).

The boundary conditions for the wave functions or envelope functions at interfaces generate the eigenvalue equations for energy; of course, different boundary conditions generate different eigenvalue equations. They are transcendental equations, involving algebraic, trigonometric, hyperbolic or even more complicated functions. With few exceptions (for instance, the Lambert equation), their solutions, which cannot be expressed as a finite combinations of elementary functions, are not systematically studied.

However, in some situations, quite accurate analytical approximate solutions can be obtained. This happens in the case of laser dressed potentials \cite{Gavrila}, \cite{Duque}, of some applications of the Floquet theory \cite{Lungu} or in the "algebraization" of some transcendental equations for eigenenergies of the bound state solutions of the Schr\"{o}dinger equation with a quite large class of potentials \cite{dABG}, \cite{VB_PM2015}. The "algebraization" consists in replacing the trigonometric functions with appropriate algebraic approximations of these functions; in this way, the transcendental eigenvalue equations can be transformed in algebraic, tractable equations. The method was introduced by de Alcantara-Bonfim and Griffiths \cite{dABG} and used in \cite{VB_PM2015} in the context of nanophysics; it generated explicit and simple solutions for the ground state energy of a particle with position independent mass (PIM) in a square well, \textit{inter alia}.

We can present now the main contributions of our paper to the aforementioned problems. Technically speaking, we shall solve the Schr\"{o}dinger equation for a particle moving in a square well, with BenDaniel - Duke boundary conditions. The interesting part of this calculation is to obtain the wave vectors (and the energies) of bound states; obtaining the form of wave functions is a trivial exercise. The wave vectors are solutions of a couple of transcendental equations, somewhat more complicated than those obtained with usual boundary conditions (corresponding to a position-independent mass (PIM)). Actually, in the PIM case, they involve only the strength of the well, $P$ - a parameter which depends both on the potential (on the depth and the length of the well) and on the mass of the particle. In the PDM case, the eigenvalue equations contain a new parameter, the ratio between the mass of the charge carrier inside and outside the well: $\beta=m_{i}/m_{o}$. These exact transcendental equations are replaced with approximate algebraic equations, as just described in the previous paragraph.

The algebraic equations obtained in this way are simpler for the ground state and the first excited states, and more complicated for higher energies. In the first case, we obtain simple and accurate expressions, for both the $\beta<1$ and $\beta>1$. The relevance of these results for type I and type II heterointerfaces are discussed.

Of course, similar issues of elementary quantum mechanical problems with BenDaniel - Duke boundary conditions have been already discussed in literature. In two recent papers, Singh and co-workers (\cite{SinghV_AJP}, \cite{SinghS_EJP}) obtained approximate solutions for the ground state wave vector replacing the trigonometric functions in the eigenvalue equation with the first few terms in their series expansions.

We consider that this paper is useful for teaching physics at tertiary level
for several reasons: (1) the problem is mathematically very simple,
essentially a Schr\"{o}dinger equation for a square potential; (2) however, it
involves a quite unfamiliar, but important concept, namely the
position-dependent mass; (3) it shows that not only wave functions, but also
enveloping functions satisfy the Schr\"{o}dinger equation; (4) it is discussed
in the framework of semiconductor nanostructures, a field of great interest
for applied and theoretical physics; (5) it uses a simple mathematical
approach to find approximate analytical solutions of transcendental
equations; (6) it illustrates the concept of type I and type II
semiconductors; (7) it suggests themes of research accessible to
undergraduate students (for instance applying the same approach to quantum
dots); (8) it makes use of a simple way of finding the real roots of a cubic
equation, a subject rarely discussed in undergraduate physics.

The structure of this paper is the following. Section II will be devoted to the
presentation of basic theory: we shall write the Schr\"{o}dinger equation for a
position-dependent mass, namely in the simplest case, when the position
dependence is given by a step function; we shall write the normalized wave functions,
inside and outside the well, for even and odd states, and the eigenvalue
equations for energy - in fact, for the wave vector. Section III contains most
of the original part of our paper. We use an algebraic approximation for $\cos
x,$ proposed by de Alcantara-Bonfim - Griffiths (dABG) \cite{dABG}, and another one, proposed by
us, for $\sin x/x;$ in this way, the eigenvalue equations become tractable,
algebraic equations - more precisely, cubic equations. We obtain simple
formulas for the wave vector of the ground state and of the first excited
state, if $\beta<1$ and if $\beta>1.$ If the limit $\beta\rightarrow1$ is
taken, in each of these cases, the results from Ref.$\ $ \cite{dABG} are obtained. We obtain
also simple formulas for shallow wells, irrespective of the parity of state
and of the value of $\beta.$ The case of excited states of higher order
$\left(  n\geq2\right)  $ is shortly discussed: the eigenvalue equation can be
approximated with a quartic algebraic equation. The results of this section are relevant for type I heterointerfaces. In Section IV, they are adapted to the case of type II heterointerfaces and to quantum dots; also, the condition of validity of the envelope function approximation are discussed.

 The last section is devoted to
conclusions. As we systematically use formulas for the real roots of a cubic
equation, where several authors propose slightly different approaches, we
collect in an appendix the minimal theoretical tools to be used in the specific
cases studied in this paper.

\section{Basic theory}

We shall solve the Schr\"{o}dinger equation for an electron moving in a square
well, described by the potential:%
\begin{equation}
V\left(  x\right)  =%
\genfrac{\{}{.}{0pt}{}{0,\ \left\vert x\right\vert \leqslant L/2}{V_{0}%
>0,\ \left\vert x\right\vert >L/2}%
\end{equation}
considering that its mass is position-dependent. More exactly, the mass inside
the well, $m_{i},$ and the mass outside the well, $m_{o},$ are different:%
\begin{equation}
m\left(  x\right)  =%
\genfrac{\{}{.}{0pt}{}{m_{i},\ \left\vert x\right\vert \leqslant
L/2}{m_{o},\ \left\vert x\right\vert >L/2}
\label{mass}
\end{equation}

\begin{figure}
  \begin{center}
  \includegraphics[width=0.5\textwidth]{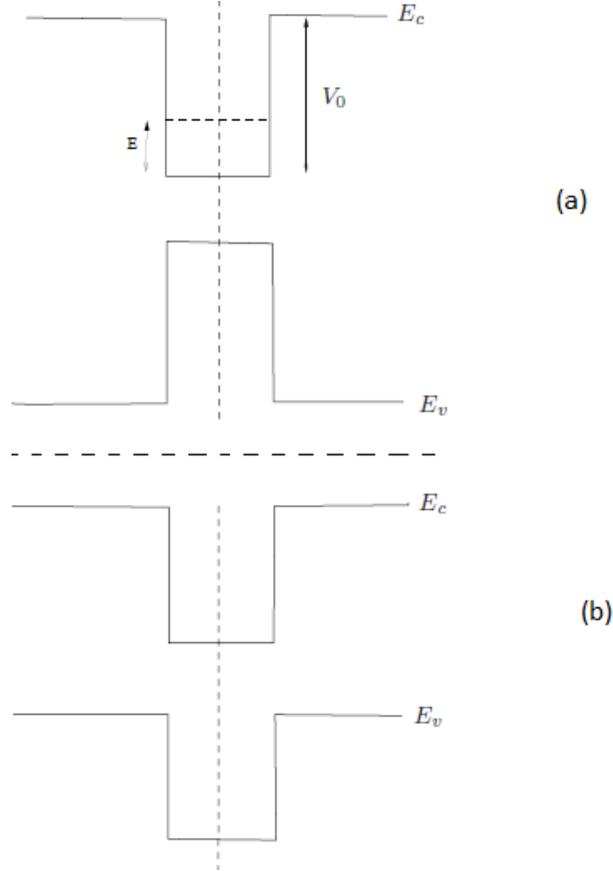}
  \end{center}
 \caption{Schematic representation of the conduction $E_c$ and valence $E_v$ bands for type I (a) and type II (b) semiconductors.}
\end{figure}

So, the Schr\"{o}dinger equation for bound states is:%
\begin{align}
\begin{aligned}
H\psi&=\left[  -\frac{\hbar^{2}}{2}\frac{d}{dx}\left(  \frac{1}{m\left(
x\right)  }\frac{d}{dx}\right)  +V\left(  x\right)  \right]  \psi_{n}
&=E_{n}\psi_{n}
\label{Schrodinger_eq}
\end{aligned}
\end{align}

Its physically acceptable solutions, i.e. the wave functions, have to satisfy
two conditions: the continuity of the wave functions and the continuity of the
probability currents density at the interface. The first one is encountered in
all quantum mechanical problems, but the second one is specific to the case of
the position dependent mass \cite{Ihn}, defined by Eq. \eqref{mass}, and takes
the form:%

\begin{equation}
\left.  \frac{1}{m_{i}}\frac{d\psi_{in}\left(  x<L/2\right)  }{dx}\right\vert
_{x\rightarrow L/2}=\left.  \frac{1}{m_{o}}\frac{d\psi_{out}\left(
x>L/2\right)  }{dx}\right\vert _{x\rightarrow L/2}
\label{Ben-DanielDuke}
\end{equation}
where $\psi_{in},\ \psi_{out}$ are the wave functions inside, respectively
outside the well. The Eq. \eqref{Ben-DanielDuke} is known as the BenDaniel - Duke boundary condition.

The $n-th$ bound state has a unique energy, $E_{n},$ but two wave vectors, one
inside the well, $k_{in,n},$ and another one outside, $k_{out,n}:$%
\begin{equation}
E_{n}=\frac{\hbar^{2}k_{in,n}^{2}}{2m_i},\ V_{0}-E_{n}=\frac{\hbar^{2}%
k_{out,n}^{2}}{2m_o}
\end{equation}
Due to the parity of the potential, $V\left(  x\right)  =V\left(  -x\right)
,$ the wave functions can be chosen to be symmetric or antisymmetric.

The symmetric wave functions, describing the even states, are:%
\begin{align}
\begin{aligned}
\psi_{2n}\left(  x,\ 0<x\leqslant L/2\right) & =A_{2n}\cos k_{in,2n}%
x\\
\psi_{2n}\left(  x,\ x>L/2\right)  &=B_{2n}\exp\left(  -k_{out,2n}x\right)
\end{aligned}
\end{align}

\begin{equation}
\psi_{2n}\left(  x<0\right)  =\psi_{2n}\left(  -x\right)
\end{equation}

The ground state wave function is, of course, $\psi_{0}\left(  x\right)  .$ The
antisymmetric wave functions, describing the odd states, are:%
\begin{align}
\begin{aligned}
\psi_{2n+1}\left(  x,\ 0<x\leqslant L/2\right)  =&A_{2n+1}\sin k_{in,2n+1}%
x\\
 \psi_{2n+1}\left(  x,\ x>L/2\right)  =&B_{2n+1,s}\exp\left(  -k_{out,2n+1}%
x\right)
\end{aligned}
\end{align}

\begin{equation}
  \psi_{2n+1}\left(  x<0\right)  =-\psi_{2n+1}\left(  -x\right)
\end{equation}

The continuity of these functions in $x=L/2$ gives:%
\begin{equation}
B_{2n}=A_{2n}\cos\frac{k_{in,2n}L}{2}\exp\left(  \frac{k_{out,2n}L}{2}\right)
\end{equation}

\begin{equation}
B_{2n+1}=A_{2n+1}\sin\frac{k_{in,2n+1}L}{2}\exp\left(  \frac{k_{out,2n+1}L}%
{2}\right)
\end{equation}
So, the wave function outside the well is:%
\begin{align}
\begin{aligned}
\psi_{2n}\left(  x>L/2\right)  =&A_{2n}\cos\frac{k_{in,2n}L}{2}
\exp\left[
-k_{out,2n}\left(  x-\frac{L}{2}\right)  \right]
\end{aligned}
\end{align}
\begin{align}
\begin{aligned}
\psi_{2n+1}&\left(  x>L/2\right)  =A_{2n+1}\sin\frac{k_{in,2n+1}L}{2}
&\exp\left[  -k_{out,2n+1}\left(  x-\frac{L}{2}\right)  \right]
\end{aligned}
\end{align}
The wave functions are normalized if:%
\begin{equation}
\frac{1}{A_{2n}^{2}}=\frac{L}{2}\left(  1+\frac{\sin k_{in,2n}L}{k_{in,2n}%
L}+\frac{1+\cos k_{in,2n}L}{k_{out,2n}L}\right)
\end{equation}

\begin{align}
\begin{aligned}
\frac{1} {A_{2n+1}^{2}}=
&\frac{L}{2}
     \left(
         1-\frac{\sin k_{in,2n+1}L} {k_{in,2n+1}L}
         +
         \frac{1-\cos k_{in,2n+1}L} {k_{out,2n+1}L}
     \right)
\end{aligned}
\end{align}
These results generalize the formula (24) in Ref.$\ $ \cite{SinghV_AJP} and the equations
(25.3e, o) in Fl\"{u}gge.\cite{Fluegge}

It is convenient to use the potential strength $P$ (introduced by Pitkanen \cite{Pitkanen},
 who actually used $\alpha,$ instead of $P$) and its inverse, $p$:
\begin{equation}
P=\sqrt{\frac{L^{2}}{2\hbar^{2}}m_{i}V_0}=\frac{1}{p}
\label{potential_strength_P}
\end{equation}
and to define also $\varepsilon_{n},\ \beta$ and $\Phi_{n}$ as:%
\begin{equation}
\varepsilon_{n}=\frac{E_{n}}{V_0}
\end{equation}

\begin{equation}
\beta=\frac{m_{i}}{m_{0}}
\end{equation}

\begin{equation}
\Phi_{n}=k_{n,in}\frac{L}{2}
\end{equation}
$P,\ \varepsilon_{n},\ \beta$ and $\Phi_{n}$ are dimensionless quantities;
$\Phi_{n}$ will be sometimes called dimensionless wave vector.

It is easy to see that:%
\begin{equation}
k_{in,n}\frac{L}{2}=P\sqrt{\varepsilon_{n}}
\end{equation}

\begin{equation}
k_{out,n}\frac{L}{2}=P\sqrt{\frac{1-\varepsilon_{n}}{\beta}}
\label{wave_vector_kout}
\end{equation}%
\begin{equation}
k_{in,n}^{2}+\beta k_{n,out}^{2}=\frac{1}{\left(  pL/2\right)  ^{2}%
}
\end{equation}

Let us mention that, if the mass is constant, i.e. if $m_{i}=m_{o}$, the eigenvalue equations are (see for instance \cite{Bastard}, p. 3. Eqs. (15), (16)):
\begin{align}
  \tan \frac{k_{in,2n}L}{2}&=\frac{k_{out}}{k_{in}}, \mathrm{\ even\ states}
    \label{Bastard_even}\\
  \tan \frac{k_{in,2n+1}L}{2}&=-\frac{k_{in}}{k_{out}}, \mathrm{\ odd\ states}
     \label{Bastard_odd}
\end{align}

If the mass is position-dependent, according to \eqref{mass}, the eigenvalue equations obtained from the Schr\"{o}dinger equation, with BenDaniel - Duke boundary conditions have the form:
\begin{align}
  \tan \frac{k_{in,2n}L}{2}&
                   =\frac{m_i}{m_o}\frac{k_{out,2n}}{k_{in,2n}}
                   =\beta\frac{k_{out,2n}}{k_{in,2n}}, \mathrm{\ even\ states}
                   \label{Ben_Dan_Duke_even}\\
   \tan \frac{k_{in,2n+1}L}{2}&=-\frac{m_o}{m_i}\frac{k_{in,2n+1}}{k_{out,2n+1}}
                          =-\frac{1}{\beta} \frac{k_{in,2n+1}}{k_{out,2n+1}}, \mathrm{\ odd\ states}
                          \label{Ben_Dan_Duke_odd}
\end{align}

We shall consider that both $m_{i}, m_{o}$ are positive; this case corresponds to type I quantum wells. So, $\beta>0$ and replacing $k_{in}L/2$ with $\Phi_{2n}$ for even states and with $\Phi_{2n+1}$ for odd states, we can put the equations \eqref{Ben_Dan_Duke_even},\eqref{Ben_Dan_Duke_odd} in a more convenient form:
\begin{align}
\begin{aligned}
\Phi_{2n}\tan\Phi_{2n}=&\frac{\sqrt{\beta}}{p}\sqrt{1-p^{2}\Phi_{2n}^{2}%
}
&, n=0,\ 1,\ ...\ \ \ \text{even states}
\label{even_states_tan}
\end{aligned}
\end{align}

\begin{align}
\begin{aligned}
\Phi_{2n+1}\cot\Phi_{2n+1}=&-\frac{\sqrt{\beta}}{p}\sqrt{1-p^{2}\Phi_{2n+1}%
^{2}}
&, n=0,\ 1,\ ...\ \ \text{odd states}
\label{odd_states_cot}
\end{aligned}
\end{align}
or, equivalently:%
\begin{align}
\begin{aligned}
\frac{\cos\Phi_{2n}}{\Phi_{2n}}=&\left(  -1\right)  ^{n}\frac{p}{\sqrt
{\beta+\left(  1-\beta\right)  p^{2}\Phi_{2n}^{2}}}
 &, n=0,\ 1,\ ...\ \ \text{even states}
\label{even_states}
\end{aligned}
\end{align}

\begin{align}
\begin{aligned}
\frac{\sin\Phi_{2n+1}}{\Phi_{2n+1}}=&\left(  -1\right)  ^{n}\frac{p}%
{\sqrt{\beta+\left(  1-\beta\right)  p^{2}\Phi_{2n+1}^{2}}}
&, n=0,\ 1,\ ...\ \text{odd states}
\label{odd_states}
\end{aligned}
\end{align}

For $\beta=1,$ they take the form of the well-known equations for the energy
eigenvalues of the finite square well. Approximate analytical solutions of
these equations were obtained for deep wells $\left(  p\ll1\right)  $ \cite{Barker}
and in the general case  \cite{dABG}, \cite{VB_PM2015},\cite{VB_RD}.

If $0<\beta<1$ ($\beta>1$), the r.h.s. of equations \eqref{even_states}, \eqref{odd_states} is a monotonically decreasing (increasing) function of $\Phi$; in both cases, the roots of these equations can be obtained using the same approach.

In this paper, we shall obtain precise analytical approximations for the
energy of the first two states, i.e. for the ground state and for the first
excited state, considering the cases $\beta<1$ and $\beta>1$ separately. For
moderate and deep wells, the formulas are both simple and accurate. In the
limit $\beta\rightarrow1,$ we shall obtain the result of de Alcantara Bonfim
and Griffiths, Eq. (17) of Ref.$\ $ \cite{dABG}. The case of shallow wells is also
discussed in detail, for both cases $\left(  \beta\lessgtr1\right)  $.

\section{Approximate analytical solutions for eigenvalue equations}

\subsection{The first even state}
According to Eq. \eqref{even_states}, the dimensionless momentum of the first even state, which
is also the ground state, is the first root of the equation:%
\begin{equation}
\frac{\cos\Phi_{0}}{\Phi_{0}}=\frac{p}{\sqrt{\beta+\left(  1-\beta\right)
p^{2}\Phi_{0}^{2}}},\ 0<\Phi_{0}<\frac{\pi}{2}
\end{equation}

We shall discuss separately the cases $\beta>1$ and $\beta<1.$

\subsubsection{The case $\beta>1$}

We shall examine this case in detail, as the approach used here will be
followed with minor changes in the other three cases (even state, $\beta<1;$
odd state, $\beta<1$ and $\beta>1$).

It is useful to introduce the new parameters $\gamma_{>}\ ,\ g_{>}%
\ ,\ A_{>}^{2}$ :
\begin{equation}
\gamma_{>}=\beta-1,\ g_{>}=\frac{1}{\gamma_{>}},\ A_{>}^{2}=\frac{P^{2}\beta
}{\beta-1}=P^{2}\beta g_{>}
\label{notations_bigger}
\end{equation}
The eigenvalue equation can be written now in a simpler form:%
\begin{equation}
\frac{\cos\Phi_{0}}{\Phi_{0}}=\frac{1}{\sqrt{\gamma_{>}}\sqrt{A_{>}^{2}%
-\Phi_{0}^{2}}},\ 0<\Phi_{0}<\frac{\pi}{2}
\label{even_states_betabigger1}
\end{equation}

In the most interesting cases, $P$ is quite large (the wells are quite deep),
and according to Eq. \eqref{notations_bigger}, $A_{>}$ is even larger, so it is more convenient to use
$A_{>}$ instead of $P$, as "large parameter".

We shall replace the exact, transcendental equation Eq. \eqref{even_states_betabigger1} with an approximate,
algebraic equation, using one of the formulas proposed in Ref.$\ $ \cite{dABG} for $\cos x$:
\begin{equation}
\cos x\simeq f\left(  x,c\right)  =\frac{1-\left(  \frac{2x}{\pi}\right)
^{2}}{\sqrt{1+cx^{2}}}
\label{dABG_formula}
\end{equation}
Ref.$\ $ \cite{dABG} gives two values for $c$, namely%
\begin{equation}
c=1-\frac{8}{\pi^{2}}=\allowbreak0.189\,43\ \ or\ \ 0.2120126
\end{equation}

\begin{figure*}
\begin{center}
\includegraphics[height=0.35\textheight]{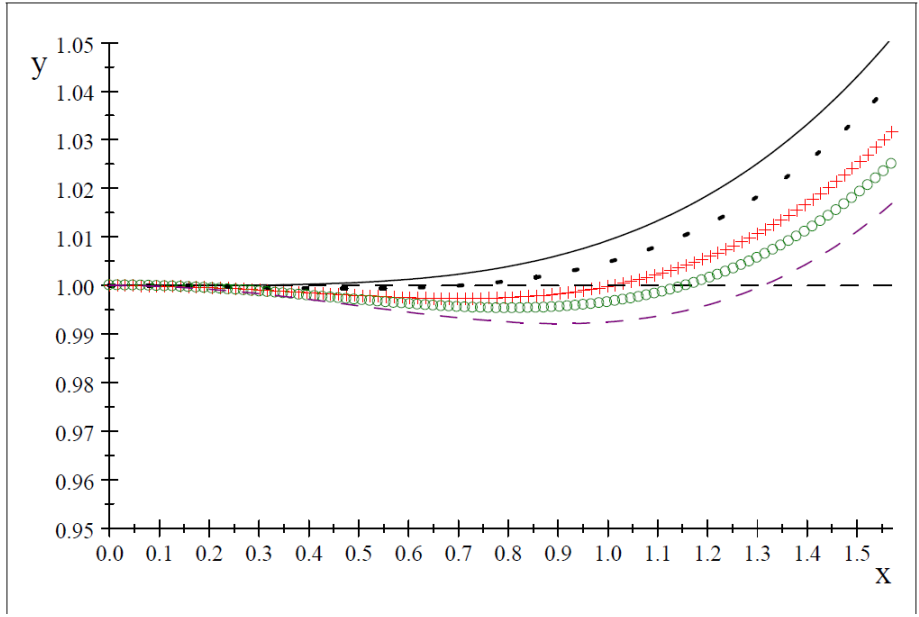}
\caption{The plots of the functions $\frac{f\left(  x,\ 0.189\,43\right)  }{\cos x}\ $ (black, solid);\ $\frac{f\left(  x,\ 0.2120126\right)  }{\cos x}$ (red,
cross);\ $\frac{f\left(  x,\ 0.2\right)  }{\cos x}$ (black, dots);\ $\frac
{f\left(  x,\ 0.22\right)  }{\cos x}$ (green, circle); $\frac{f\left(
x,\ 0.23\right)  }{\cos x}$ (purple, dash); $y\left(  x\right)  =1,$ (black, dash)}
\label{errors_cos}
\end{center}
\end{figure*}

The relative errors of the algebraic approximation of $\cos$ are plotted in
Fig. \ref{errors_cos}. For $0<x\lesssim0.7,$ the both approximations are excellent, with
relative errors of about $10^{-3}$, but for $0.7<x\lesssim\pi/2$, the second
one is better; even in this case, the error increases until about $3.7\%$ for
$x\lesssim\pi/2.\,$\ So, the value $c=0.2120126$ seems to be more appropriate.
In the same time, larger values of $c,$ for instance $c=0.22$ or $c=0.23,$
give better precision at larger values of $x.$ We shall keep the value of $c$
undefined in our formulas for the roots; eventually, it can be adapted
according to the value of the specific root. The advantage of using a very
precise value of $c$ is not really obvious.%

So, we get for the algebraic approximation of the eigenvalue equation:%
\begin{equation}
\frac{1}{\Phi_{0}}\frac{1-\left(  \frac{2\Phi_{0}}{\pi}\right)  ^{2}}%
{\sqrt{1+c\Phi_{0}^{2}}}=\frac{1}{\sqrt{\gamma_{>}}\sqrt{A_{>}^{2}-\Phi
_{0}^{2}}}
\end{equation}
With
\begin{equation}
\Phi_{0}^{2}=z
\end{equation}
it can be written as:%
\begin{align}
\begin{aligned}
z^{3}+\left(  \frac{1}{16}\pi^{4}cg_{>}-A_{>}^{2}-\frac{1}{2}\pi^{2}\right)
z^{2}+
\frac{\pi^{2}}{2}\left(  \frac{\pi^{2}}{8}g_{>}+A_{>}^{2}+
+\frac{1}{8}%
\pi^{2}\right)z
&
-\frac{1}{16}\pi^{4}A_{>}^{2}=0
\label{cubic_eq_even_bigger}
\end{aligned}
\end{align}

Following the approach outlined in the Appendix, let us put:%
\begin{equation}
a_{2>}=\left(  \frac{1}{16}\pi^{4}cg_{>}-A_{>}^{2}-\frac{1}{2}\pi^{2}\right)
\label{a2_bigger_init}
\end{equation}

\begin{equation}
a_{1>}=\frac{\pi^{2}}{2}\left(  \frac{\pi^{2}}{8}g_{>}+A_{>}^{2}+\frac{1}%
{8}\pi^{2}\right)
\label{a1_bigger_init}
\end{equation}

\begin{equation}
a_{0>}=\left(  -\frac{1}{16}\pi^{4}A_{>}^{2}\right)
\label{a0_bigger_init}
\end{equation}

\begin{equation}
p_{c>}=-\frac{1}{3}\left(  A_{>}^{4}+P_{1>}A_{>}^{2}+P_{0>}\right)
\label{pc_bigger}
\end{equation}
where $P_{1>}$ is a linear, and $P_{0>}$ a quadratic polynomial in $g_{>}$.

In the last equation, the index $c$ in $p_{c>}$ is for $\cos$, suggesting the
even symmetry of this state.

If the well is not too shallow, $A_{>}>1$ and it
is convenient to define the "small" parameter $\alpha_{>}$:%
\begin{equation}
\alpha_{>}=\frac{1}{A_{>}}
\end{equation}
so we can write:%
\begin{equation}
p_{c>}=-\frac{A_{>}^{4}}{3}\left(  1+P_{1>}\alpha_{>}^{2}+P_{0>}\alpha_{>}%
^{4}\right)
\end{equation}

Similarly, defining $Q_{K>}$ through the formula:%
\begin{align}
\begin{aligned}
q_{c>}&=\frac{2}{27}A_{>}^{6}Q_{K>}
&=\frac{2}{27}A_{>}^{6}\left(  1+Q_{2>}%
\alpha_{>}^{2}+Q_{1>}\alpha_{>}^{4}+Q_{0>}\alpha_{>}^{6}\right)
\end{aligned}
\end{align}
where:%
\begin{equation}
Q_{2>}=-\frac{3\pi^{2}}{4}\left(  1+\frac{1}{4}\pi^{2}cg_{>}\right)
\end{equation}

\begin{align}
\begin{aligned}
Q_{1>}=\frac{3^{2}\pi^{4}}{2^{4}}
   \left[ \frac{\pi^{4}}{2^{4}\cdot3}c^{2}g_{>}^{2}
         -
           \frac{1}{2}
           \left(
             \frac{\pi^{2}}{2\cdot3}c+1
           \right)
           g_{>}
         +
         \frac{1}{3}
    \right]
\end{aligned}
\end{align}

\begin{align}
\begin{aligned}
Q_{0>}=
       -\frac{\pi^{6}}{2^{6}}
        \left[
           \frac{\pi^{6}}{2^{6}}
           c^3g_>^3
           -
           \frac{3^2\pi^2c}{2^3}
           \left(
             \frac{\pi^2c}{3}
             +
             1
           \right)
           g_>^2
         +
          3^2
          \left(
            \frac{5\pi^2c}{2^3\cdot3}
            +
            1
          \right)
          g_>
         +
          1
        \right]
\end{aligned}
\end{align}
we obtain the following expression for $K_{c>}:$%
\begin{align}
\begin{aligned}
K_{c>}&=\frac{Q_{K>}}{\left\vert P_{K>}\right\vert ^{3/2}}
&=
\frac{1+Q_{2>}%
 \alpha_{>}^{2}+Q_{1>}\alpha_{>}^{4}+Q_{0>}\alpha_{>}^{6}}{\left\vert 1+P_{1>}\alpha
_{>}^{2}+P_{0>}\alpha_{>}^{4}\right\vert ^{3/2}}
\label{Kc_bigger}
\end{aligned}
\end{align}

In fact, the index $K$ in $Q_{K>}$ was chosen because $Q_{K>}$ is the part of
$q_{c>}$ which enters effectively in the simplest form of $K_{c>}.\ $For deep wells, $\alpha_>
\rightarrow0$; consequently $\ p_{c}<0$ and:%
\begin{equation}
K_{c>}=\frac{1+Q_{2>}\alpha_{>}^{2}+Q_{1>}\alpha_{>}^{4}+Q_{0>}\alpha_{>}^{6}}
{\left(
1+P_{1>}\alpha_{>}^{2}+P_{0>}\alpha_{>}^{4}\right)  ^{3/2}}
\end{equation}

It is fortunate, for the precision and simplicity of our formulas, that the
quadratic and quartic terms of its series expansion in $\alpha_>$ are zero, and%
\begin{equation}
K_{c>}=1-\frac{3^3\pi^{6}}{2^7}g_{>}\left(  1+\frac{\pi^{2}}{4}c\right)
\alpha_{>}^{6}+...
\end{equation}

The equation has three real roots; we shall choose the root which is
compatible with the condition $0<\Phi<\pi/2,$ or $z<\pi^{2}/4.$ Putting:%
\begin{equation}
K_{c>}=1-Z_{>}\ ,\ Z_{>}\ll1
\end{equation}
we find easily:%
\begin{align}
\begin{aligned}
y_{1}&=\sin\left(  \frac{1}{3}\arcsin\left(  1-Z\right)  -\frac{\pi}{3}\right)
&
=-\frac{1}{2}-\sqrt{\frac{Z}{6}}+...
\end{aligned}
\end{align}

\begin{equation}
y_{2}=\sin\left(  \frac{1}{3}\arcsin\left(  1-Z\right)  +\frac{\pi}{3}\right)
=1-\frac{Z}{9}+...
\end{equation}

\begin{equation}
y_{3}=-\sin\left(  \frac{1}{3}\arcsin\left(  1-Z\right)  \right)  =-\frac
{1}{2}+\sqrt{\frac{Z}{6}}+...
\end{equation}
Finally, using the formula Eq. \eqref{sol_cubic_eq}:%
\begin{equation}
z_1=\sqrt{\frac{4\left\vert p_{c>}\right\vert }{3}}y_1-\frac{1}{3}%
a_{2>}
\end{equation}
and introducing the notation:%
\begin{equation}
C=\frac{\pi^{2}}{2}c
\end{equation}
we get (later on we shall drop the index $1$ of $z_{1}$):%

\begin{align}
 \begin{aligned}
z&\left(  \beta>1\right)  =\frac{\pi^{2}}{4}-
 \frac{\pi^{3}}{8}
 \sqrt{  1+\frac{C}{2}}
 \left(  \sqrt{g_{>}}\alpha_{>}\right)
 +\frac{\pi^{4}}{32}\left(  1+C\right)
  \left(  \sqrt{g_{>}}\alpha_{>}\right)^{2}
 \\
& +\frac{\pi^{5}}{32}
   \left(1+\frac{Cg_{>}}{2}\right)
   \sqrt{  1+\frac{C}{2}}\sqrt{g_{>}%
}\alpha_{>}^{3}
+\frac{\pi^{6}}{128}g_{>}\left(  1+C\right)  \left(
1+\frac{Cg_{>}}{2}\right)  \alpha_{>}^{4}+...
\label{sol_approx_even_bigger}
 \end{aligned}
\end{align}

If the depth of the well increases indefinitely, $\alpha_{>}\rightarrow0$ and
$z_{1}\rightarrow\pi^{2}/4,\ \Phi_{0}\rightarrow\pi/2,$ as requested. Indeed, in a
finite well, the energy of a bound state is smaller than the corresponding
energy in an infinite one, so the first term in $\sqrt{g_{>}}\alpha_{>}$ in
the previous formula is negative. \qquad\qquad

It is useful to write \eqref{sol_approx_even_bigger} in terms of more physical parameters, $p$ and $\beta$. In order to do
this, let us notice that:%
\begin{equation}
g_{>}\alpha_{>}^{2}=\frac{p^{2}}{\beta},\ \ \alpha_{>}^{2}=\frac{\beta
-1}{\beta}p^{2}
\end{equation}
so Eq. \eqref{sol_approx_even_bigger} takes the form:
\begin{align}
\begin{aligned}
z\left(  \beta>1\right)=&\frac{\pi^{2}}{4}
                        -\frac{\pi^{3}}{8}
                        \sqrt{  1+\frac{C}{2}}
                        \frac{p}{\beta^{1/2}}
                        +\frac{\pi^{4}}{32}\left(  1+C\right)
                         \frac{p^{2}}{\beta}+ \\
                        &+\frac{\pi^{5}}{32}
                          \sqrt{1+\frac{C}{2}}
                          \left(
                           \beta+\frac{C}{2}-1
                          \right)
                          \frac{p^{3}}{\beta^{3/2}}
+
\frac{\pi^{6}}{128}
\left(  1+C\right)
\left(  \beta+\frac{C}{2}-1 \right)
\frac{p^{4}}{\beta^{2}}+...
\label{sol_approx_even_bigger_2}
\end{aligned}
\end{align}
and:
\begin{align}
\begin{aligned}
z\left(  \beta=1\right)
=&\frac{\pi^{2}}{4}
-
\frac{\pi^{3}}{8}
\sqrt
{
  1+\frac{C}{2}
}
p
+\frac{\pi^{4}}{32}\left(  1+C\right)  p^{2}\\
&+\frac
{\pi^{5}}{32}\sqrt{1+\frac{C}{2}}\frac{C}{2}p^{3}
+
\frac{\pi^{6}}{128}\frac
{C}{2}\left(  1+C\right)  p^{4}+...
\end{aligned}
\label{sol_approx_even_equal}
\end{align}

It is a simple exercise to check that the previous formula coincides with the
first three terms of the power series given by Eq. (17) of Ref.$\ $ \cite{dABG}.

For the other two roots given by Eq. \eqref{sol_cubic_eq}, $z_{2},\ z_{3}$  the "large" parameter $A_{>}$ does
not disappear, so the condition $z^{2}<\pi^{2}/4$ cannot be fulfilled.

If the parameter $\alpha_{>}$ cannot be considered "small", the formulas \eqref{pc_bigger}
and \eqref{Kc_bigger}, used according to the algorithm given in the Appendix, will give the
root of the cubic equation \eqref{cubic_eq_even_bigger}.

\subsubsection{The case $\beta<1$}

If$\ \beta<1$, the eigenvalue equation for the dimensionless wave vector is:%
\begin{equation}
\frac{\cos\Phi_{0}}{\Phi_{0}}=\frac{1}{\sqrt{\gamma_{<}}\sqrt{A_{<}^{2}%
+\Phi_{0}^{2}}},\ \ 0<\Phi_{0}<\frac{\pi}{2}
\end{equation}
with the following definitions for the parameters:%
\begin{equation}
\gamma_{<}=1-\beta,\ \frac{1}{\gamma_{<}}=g_{<},\ A_{<}^{2}=\frac{P^{2}\beta
}{1-\beta}=P^{2}\beta\gamma_{<}
\end{equation}

Using the dABG algebraization for $\cos \Phi_0,$ we get the algebraic equation:%
\begin{equation}
\frac{1}{\Phi_0}\frac{1-\left(  \frac{2\Phi_0}{\pi}\right)  ^{2}}{\sqrt{1+c\Phi_0^{2}}%
}=\frac{1}{\sqrt{\gamma_{<}}\sqrt{A_{<}^{2}+\Phi_0^{2}}}
\end{equation}
It becomes, with%
\begin{equation}
\Phi_{0}^{2}=z
\end{equation}
a cubic equation:%
\begin{equation}
z^{3}+\left(  A_{<}^{2}-\frac{1}{2}\pi^{2}-\frac{1}{16}\pi^{4}cg_{<}\right)
z^{2}+\frac{\pi^{2}}{2}\left(  \frac{1}{8}\pi^{2}-A_{<}^{2}-\frac{\pi^{2}}%
{8}g_{<}\right)  \allowbreak z+\frac{1}{16}\pi^{4}A_{<}^{2}=0,\ \ \ \ \beta
<1
\label{cubic_eq_even_smaller}
\end{equation}

Similar to the case $\beta>1,$ let us put:%
\begin{equation}
a_{2<}\left(  A_{<}^{2},\ g_{<}\right)  = A_{<}^{2}-\frac{1}{2}\pi
^{2}-\frac{1}{16}\pi^{4}cg_{<}
\end{equation}

\begin{equation}
a_{1<}\left(  A_{<}^{2},\ g_{<}\right)  =\frac{\pi^{2}}{2}\left(  \frac{1}%
{8}\pi^{2}-A_{<}^{2}-\frac{\pi^{2}}{8}g_{<}\right)
\end{equation}

\begin{equation}
a_{0<}\left(  A_{<}^{2},\ g_{<}\right)  =\frac{1}{16}\pi^{4}A_{<}%
^{2}
\end{equation}
Let us notice that:%
\begin{align}
\begin{aligned}
a_{2<}\left(
         A_{<}^{2},
         \ g_{<}
       \right) & =
       a_{2>}
       \left(
          A_{>}^{2}\rightarrow-A_{<}^{2},g_{>}\rightarrow-g_{<}\right)
\\
 a_{1<}\left(
A_{<}^{2},\ g_{<}\right)  &=a_{1>}\left(  A_{>}^{2}\rightarrow-A_{<}%
^{2},g_{>}\rightarrow-g_{<}\right)
\\
 a_{0<}\left(  A_{<}^{2}\right)
&=a_{0>}\left(
     A_{>}^{2}\rightarrow-A_{<}^{2}\right)
\label{schimbare_de_semn}
\end{aligned}
\end{align}
As in Eq. \eqref{sol_approx_even_bigger}, the parameters $g_{>}$, $\alpha_{>}$ enter only through the monoms $g_{>}\alpha_{>}^2$ and $\alpha_{>}^2$ at various powers.  The roots of Eq. \eqref{cubic_eq_even_smaller} can be obtained from Eq. \eqref{sol_approx_even_bigger} making the
following substitutions:
\begin{equation}
g_{>}\rightarrow-g_{<},\ \alpha_{>}^{2}\rightarrow-\alpha_{<}^{2}
\end{equation}
The result is:%
\begin{align}
\begin{aligned}
z&\left(  \beta<1\right) =
\frac{\pi^{2}}{4}-\frac{1}{8}\pi^{3}\sqrt
{g_{<}\left(  1+\frac{C}{2}\right)  }\alpha_{<}
+
\frac{1}{32}\pi^{4}%
g_{<}\left(  1+C\right)  \alpha_{<}^{2}+\\
&
+\frac{\pi^{5}}{32}\left(
1-\frac{Cg_{<}}{2}\right)  \sqrt{g_{<}\left(  1+\frac{C}{2}\right)  }%
\alpha_{<}^{3}
-
\frac{\pi^{6}}{128}g_{<}\left(  1+C\right)  \left(  1-\frac
{C}{2}g_{<}\right)  \alpha_{<}^{4}+...
\label{sol_approx_even_smaller_2}
\end{aligned}
\end{align}

In order to write the previous equation in terms of $p$ and $\beta$, let us notice, similarly to
the previous case, that%
\begin{equation}
\sqrt{g_{<}}\alpha_{<}=\frac{p}{\sqrt{\beta}},\ \alpha_{<}^{2}=p^{2}%
\frac{1-\beta}{\beta}
\end{equation}
Replacing these monoms in \eqref{schimbare_de_semn}, we obtain an equation having exactly the form \eqref{sol_approx_even_smaller_2}.
\begin{align}
\begin{aligned}
z\left( p, \beta<1\right)
=&
  \frac{\pi^{2}}{4}
  -
   \frac{\pi^{3}}{8}
   \sqrt{\frac{C}{2}+1}
   \frac{p}{\sqrt{\beta}}
  +
  \frac{1}{32}\pi^{4}\left( C+1 \right)  \frac{p^{2}}{\beta} \\
  &
  -
  \frac{\pi^{5}}{32}
  \sqrt{\left( \frac{C}{2}+1 \right)}
  \left(  \beta+\frac{C}{2}-1 \right)
  \frac{p^3}{\beta^{3/2}}
 +
  \frac{\pi^{6}}{128}
  \left( C+1 \right)
  \left( \beta+\frac{C}{2}-1 \right)
  \frac{p^{4}}{\beta^{2}}+...
\label{sol_approx_even_smaller}
\end{aligned}
\end{align}

\subsection{The case of a shallow well}

It is well-known that, for $\beta=1,$ any square well, even a very shallow
one, has at least one bound state; if the well keeps only one state, this state is an even one. We shall see that this property remains
valid if $\beta\neq1.$

A shallow well corresponds to small values of $P$\ and large values of $p.$ In
this case, replacing in the eigenvalue equation the zero order approximation
for $\cos\Phi_{0},$ i.e. $\cos\Phi_{0}\simeq1,$ one obtains%
\begin{equation}
\frac{1}{\Phi_{0}^{2}}=\frac{p^{2}}{\beta+\left(  1-\beta\right)  p^{2}%
\Phi_{0}^{2}}
\end{equation}
Its solution is independent of $\beta$:%
\begin{equation}
\Phi_{0}=\frac{1}{p}=P
\end{equation}
The relative error of this solution is about $-10^{-2}$ for $P=0.1$ and
becomes about $-3.7\times10^{-2}$ for $P=0.2.$

A more precise expression for the root can be obtained using the parabolic
approximation, i.e. $\cos \Phi_{0}\simeq1-\frac{\Phi_{0}^{2}}{2},$ gives:%
\begin{equation}
\frac{\cos^{2}\Phi_{0}}{\Phi_{0}^{2}}=\frac{\left(  1-\Phi_{0}^{2}/2\right)
^{2}}{\Phi_{0}^{2}}=\frac{p^{2}}{\beta+\left(  1-\beta\right)  p^{2}\Phi
_{0}^{2}}
\end{equation}
Putting:%
\begin{equation}
\Phi_{0}^{2}=z,\ \ 1-\beta=\frac{1}{\gamma},\ \beta\gamma=\delta=\frac{\beta
}{1-\beta}
\end{equation}
the equation can be written as:%
\begin{equation}
z^{3}+\left(  \delta P^{2}-4\right)  z^{2}-4\delta\left(  1+P^{2}\right)
z+4\delta P^{2}=0
\end{equation}
We get:%
\begin{equation}
p_{c}=-\frac{16}{3}\left(  1+\frac{3}{4}\beta\gamma+\frac{1}{4}P^{2}%
\beta\gamma+\frac{1}{16}\left(  P^{2}\beta\gamma\right)  ^{2}\right)
\end{equation}

\begin{align}
\begin{aligned}
q_{c}=&
     \frac{128}{27}
     \left(  1+\frac{9}{4}\beta\gamma
             -\frac{15}{32}P^{2}\beta\gamma
             -\frac{9}{32}\left(  \beta\gamma P\right)^{2}
     \right)-\\
     &-
     \frac{128}{27}
     \left(
          \frac{3}{32}\left(  P^{2}\beta\gamma\right)^{2}
         +\frac{1}{64}\left(P^{2}\beta\gamma\right)^{3}
     \right)
\end{aligned}
\end{align}

\begin{equation}
K=\frac{1+\frac{9}{4}\allowbreak\beta\gamma-\frac{15}{32}P^{2}\beta
\gamma-\frac{9}{32}\left(  \beta\gamma P\right)  ^{2}-\frac{3}{32}\left(
P^{2}\beta\gamma\right)  ^{2}-\frac{1}{64}\left(  P^{2}\beta\gamma\right)
^{3}}{\left\vert 1+\frac{3}{4}\beta\gamma+\frac{1}{4}P^{2}\beta\gamma+\frac
{1}{16}\left(  P^{2}\beta\gamma\right)  ^{2}\right\vert ^{3/2}}
\label{K_shallow_well}
\end{equation}

As $P\ll1,\ K$ is well approximated by:%
\begin{align}
\begin{aligned}
K=\sqrt{d}\left(  1+\frac{9}{4}\beta\gamma\right)
+\frac{3}{32}\beta
\gamma\left(  4-5\sqrt{d}+9\beta\gamma\right)  P^{2}%
+...
\label{K_approx_shallow_well}
\end{aligned}
\end{align}
where we used the notation:%
\begin{equation}
d=\sqrt{\frac{3}{4}\beta\gamma+1}
\end{equation}
and presumed that the physically reasonable condition:%
\begin{equation}
\frac{3}{4}\beta\gamma+1>0
\label{condition}
\end{equation}
The roots can be easily obtained, using the same method used in the previous
subsections. If the condition \eqref{condition} is not satisfied, and $p_{c}\left(
P=0\right)  >0,$ the expression under modulus in Eq. \eqref{K_shallow_well} changes its sign and a
similar series expansion for $K$ can be obtained.

\subsection{{The first odd state}}

\subsubsection{The case $\beta>1$}

The exact eigenvalue equation for the first odd state - which is also the
first excited state - can be written as:%
\begin{equation}
\frac{\sin\Phi_{1}}{\Phi_{1}}=\frac{1}{\sqrt{\gamma_{>}}\sqrt{A_{>}^{2}%
-\Phi_{1}^{2}}},\ \frac{\pi}{2}<\Phi_{1}<\pi
\label{odd_equation}
\end{equation}

As the shape of the function $\sin \Phi_{1}/\Phi_{1}$ on the interval $\left[  0,\pi\right]
$ is quite similar with the shape of $\cos \Phi_{1}$ on the interval $\left[
0,\pi/2\right]  ,$ we can try an algebraization for $\sin \Phi_{1}/\Phi_{1}$ similar to that
proposed by de Alcantara Bonfim and Griffiths \cite{dABG} for $\cos \Phi_{1}:$%
\begin{equation}
\frac{\sin\Phi_{1}}{\Phi_{1}}\simeq\frac{1-\left(  \Phi_{1}/\pi\right)  ^{2}%
}{\sqrt{1+a\Phi_{1}^{2}}}
\label{sinc_approx}
\end{equation}

\begin{figure*}
\begin{center}
\includegraphics[height=0.35\textheight]{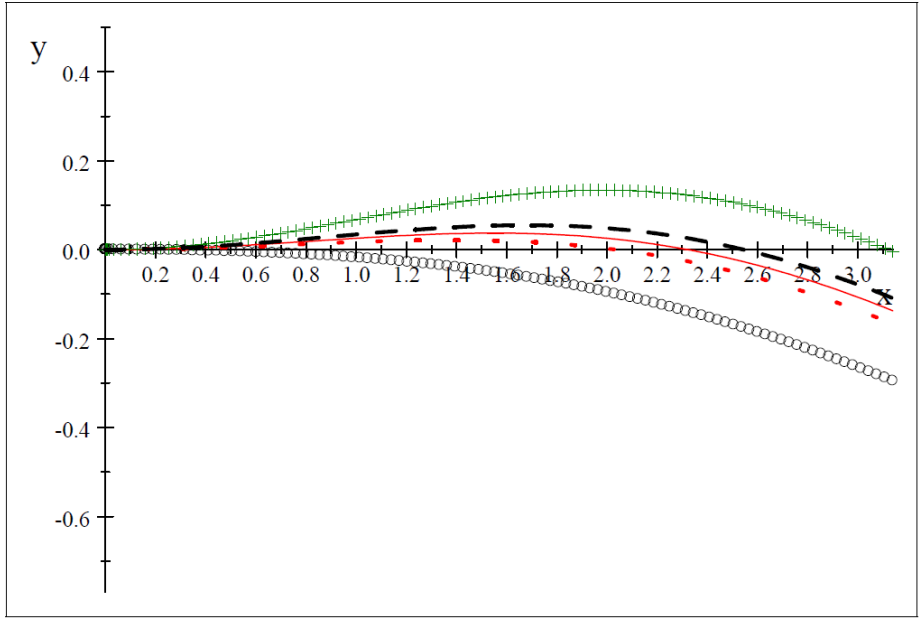}
\caption{The plot of the function $F\left(  a;x\right)  ,$ Eq. \eqref{F_function}; $a=0.1$ (black,
circle); $0.18$ (red, dots) $0.2$ (red, solid); $0.22$ (black, dash); $0.3$
(green,cross)}
\label{errors_sin}
\end{center}
\end{figure*}

Plotting the function%
\begin{equation}
F\left(  a;x\right)  =\frac{\sin x}{x}\frac{\sqrt{1+ax^{2}}}{1-\left(
x/a\right)  ^{2}}-1
\label{F_function}
\end{equation}
for several values of $a,$ we can see that the error of this approximation is
of the order $10^{-2}$ and varies quite strongly (for the same value of $a $)
if $x$ moves on the interval $\left[  0,\pi\right]  ;$ so, fixing $a$ to a very precise
value seems a procedure of questionable usefulness. Let us mention again that
the physically interesting root of Eq. \eqref{sinc_approx} belongs to the interval $\left(
\pi/2,\pi\right)$ \cite{VB_PM2015}.  More precise could be an iterative process, namely
obtaining a root $x\left(  a\right)  $ for a fixed $a$ and repeating the
calculation of this root using a new value, $a_{0},$ given by graphical
inspection of the condition $F\left(  a_{0};x\left(  a\right)  \right)
\simeq0$ in Fig. \ref{errors_sin}. In our numerical verifications, we chose the value
$a=0.2.$%

So, replacing Eq. \eqref{sinc_approx} in the l.h.s. of the Eq. \eqref{odd_equation}, and putting:%
\begin{equation}
\Phi_1^{2}=z
\end{equation}
we obtain a cubic equation:%
\begin{align}
\begin{aligned}
z^{3}-\left(  A_{>}^{2}+2\pi^{2}\right)  z^{2}+\pi^{2}\left(  2A_{>}^{2}%
+\pi^{2}+\pi^{2}ag_{>}\right)  z
+\pi^{4}\left(  g_{>}-A_{>}
^{2}\right)  =0
\end{aligned}
\end{align}
Let us define:%
\begin{equation}
a_{2>}=-\left(  A_{>}^{2}+2\pi^{2}\right)
\label{a2_bigger}
\end{equation}

\begin{equation}
a_{1>}=\pi^{2}\left(  2A_{>}^{2}+\pi^{2}+\pi^{2}ag_{>}\right)
\label{a1_bigger}
\end{equation}

\begin{equation}
a_{0>}=\allowbreak\pi^{4}\left(  g_{>}-A_{>}^{2}\right)
\label{a0_bigger}
\end{equation}

In order to avoid too complicated notations, we used in Eqs. \eqref{a2_bigger},\eqref{a1_bigger},\eqref{a0_bigger} the same
symbol as in Eqs. \eqref{a2_bigger_init},\eqref{a1_bigger_init},\eqref{a0_bigger_init} in the spirit of the definition \eqref{cubic_eq_z} in the
Appendix; we hope that this ambiguity will not produce any confusions.
Following exactly the same steps as for even states, we find (the index $s$ is
for $\sin$, suggesting an odd state):
\begin{equation}
p_{s>}=-\frac{1}{3}A_{>}^{4}\left(  1-2\pi^{2}\alpha_{>}^{2}-\pi^{4}\left(
3ag_{>}-1\right)  \alpha_{>}^{4}\right)
\end{equation}

\begin{align}
\begin{aligned}
q_{s>}&=
     \frac{2}{27}A_{>}^{6}
     \left[  1-3\pi^{2}\alpha_{>}^{2}+
          3\pi^{4}
          \left(
              1-\frac{3}{2}ag_{>}
          \right)
          \alpha_{>}^{4}
        -
          \pi^{4}
          \left(
             \frac{27}{2}g_{>}
             +\pi^{2}+9\pi^{2}ag_{>}
          \right)\alpha_{>}^{6}
     \right]
\end{aligned}
\end{align}

\begin{equation}
K_{s>}=\frac{\left(  1-3\pi^{2}\alpha_{>}^{2}+3\pi^{4}\left(  1-\frac{3}%
{2}ag_{>}\right)  \alpha_{>}^{4}-\pi^{4}\left(  \frac{27}{2}g_{>}+\pi^{2}%
+9\pi^{2}ag_{>}\right)  \alpha^{6}\right)  }{\left\vert \allowbreak1-2\pi
^{2}\alpha_{>}^{2}-\pi^{4}\left(  3ag_{>}-1\right)  \alpha_{>}^{4}\right\vert
^{3/2}}
\end{equation}

If%
\begin{equation}
\ A_{>}=P\sqrt{\frac{\beta}{\beta-1}}>1\Leftrightarrow\ P>\sqrt{\frac{\beta
-1}{\beta}}
\end{equation}
the power expansion for $K_{>}$ is:%
\begin{equation}
K_{s>}=\allowbreak1-\frac{27}{2}\pi^{4}g_{>}\left(  \pi^{2}a+1\right)
\allowbreak\alpha_{>}^{6}+\allowbreak O\left(  \alpha^{7}_>\right)
\end{equation}
and the physically convenient root has the form:%

\begin{equation}
z\left(  \beta>1\right)  =\pi^{2}-\pi^{2}\sqrt{g_{>}\left(  1+\pi
^{2}a\right)  }\alpha_{>}+\frac{\pi^{4}}{2}ag\allowbreak_{>}\alpha_{>}^{2}%
+\pi^{4}\sqrt{g_{>}\left(  1+\pi^{2}a\right)  }\alpha_{>}^{3}+\frac{\pi^{6}%
}{2}ag_{>}\alpha_{>}^{4}+...
\label{sol_approx_odd_bigger}
\end{equation}

In terms of $p$ and $\beta:$%
\begin{align}
\begin{aligned}
z\left(  \beta>1\right)
 =&
\pi^{2}-\pi^{2}\sqrt{  1+\pi^{2}a}
        \frac{p}{\sqrt{\beta}}
+\frac{\pi^{4}}{2}a\frac{p^{2}}{\beta}
+\pi^{4}%
\sqrt{1+\pi^{2}a}\left(  \beta-1\right)
\left(  \frac
          {p}{\sqrt{\beta}}
\right)^{3}+
\\
&
+
\frac{\pi^{6}}{2}a\left(  \beta-1\right)
\frac{p^{4}}{\beta^{2}}+...
\label{sol_approx_odd_bigger_beta}
\end{aligned}
\end{align}

\subsubsection{The case $\beta<1$}

The exact eigenvalue equation for the first odd state, if $\beta<1,$ can be
written as:%
\begin{equation}
\frac{\sin \Phi_1}{\Phi_1}=\frac{1}{\sqrt{\gamma_{<}}\sqrt{A_{<}^{2}+\Phi_1^{2}}}%
,\ 0<\Phi_1<\pi
\end{equation}
With the algebraic approximation for $\sin \Phi_1/\Phi_1$, Eq. \eqref{sinc_approx}, it gives the following
cubic equation in $z=\Phi_1^{2}:$%
\begin{align}
\begin{aligned}
z^{3}+\left(  A_{<}^{2}-2\pi^{2}\right)  z^{2}
+\left(
       -2\pi^{2}A_{<}^{2}%
      +\pi^{4}-\pi^{4}ag_{<}
\right)z
+
\left(
  \pi^{4}A_{<}^{2}-
  \pi^{4}g_{<}
\right)  =0
\end{aligned}
\end{align}

Following the same steps as before, we find:%
\begin{align}
\begin{aligned}
z\left(  \beta<1\right)  =
       \pi^{2}-
       \pi^{2}
       \sqrt{
         g_{<}
         \left(  1+\pi^{2}a
         \right)
           }
         \alpha_{<}
       +
         \frac{\pi^{4}}{2}ag_{<}
         \alpha_{<}^{2}
       -
         \pi^{4}
         \sqrt{g_{<}
             \left(  1+\pi^{2}a\right)
              }
         \alpha_{<}^{3}
         -
         \frac{\pi^{6}}
              {2}ag_{<}\alpha_{<}^{4}+...
\end{aligned}
\end{align}

For both cases - $\beta\lessgtr1$ - in the limit of an infinitely deep well,
$z\left(  \beta\lessgtr1,\ \alpha_{<}=0\right)  =\pi^{2},\ \Phi_{1}\left(
\beta\lessgtr1,\ \alpha_{<}=0\right)  =\pi,$ as requested, and the first
correction to this value is negative. The expression of the root in terms of $p$ and $\beta$
is identical with \eqref{sol_approx_odd_bigger_beta}.

  The relative errors of the formulas \eqref{sol_approx_even_bigger_2}, \eqref{sol_approx_even_smaller}, \eqref{sol_approx_odd_bigger}, with respect to the exact roots of the corresponding algebraic equations, are very small - of about $10^{-4}...10^{-6}$ for physically interesting values of the parameters $p, \beta, a, c$. The order of magnitude of the errors introduced by the algebraization of transcendental equations of the eigenvalues can be visualized in Fig. \ref{errors_cos} and \ref{errors_sin}. In other words, the main contribution to the errors of our results is given by the approximation of trigonometric functions with algebraic ones, not by the approximation of the exact formulas of the roots of cubic equations with the low order terms of their series expansions. As already mentioned, one of the physical motivations of the calculation of the energy of bound states in heterostructures is to explain their photoluminescence properties. In several cases (see for instance Ref.$\ $ \cite{Biswas}), the authors use Barker's formula for the energy levels in a square well \cite{Barker}. Much more precise analytical expressions for these energy are available in the literature \cite{dABG}, \cite{VB_PM2015}, for the case of constant mass; in this paper, we propose similar formulas, considering the case of position-dependent mass.

\subsection{Higher order states}

In the previous subsections, we analyzed the ground state $\left(  n=0\right)
$\ and the first excited state $\left(  n=1\right)  $ of a square well, with
BenDaniel - Duke boundary conditions. For $n\geqslant2,$ the de Alcantara -
Bonfim formula \eqref{dABG_formula} can be extended to larger arguments:%
\begin{align}
\begin{aligned}
\cos\Phi\simeq\frac{1-4\left(  \Phi-2n\pi\right)  ^{2}/\pi^{2}}{\sqrt
{1+c\left(  \Phi-2n\pi\right)  ^{2}}},\ 2n\pi<\Phi<2n\pi+\frac{\pi}%
{2}
\end{aligned}
\end{align}
but the eigenvalue equation, obtained in this way, is a sextic equation (which
cannot be reduced to a cubic equation in $\Phi^{2}$), so cannot be solved. We
meet similar difficulties if we try to use the eigenvalue equation in terms of
the $\tan$ function, replaced by the approximate expression  \cite{VB_PM2015}:

\begin{figure}[h!t]
  \centering
  \subfigure[]{\includegraphics[width=\textwidth]{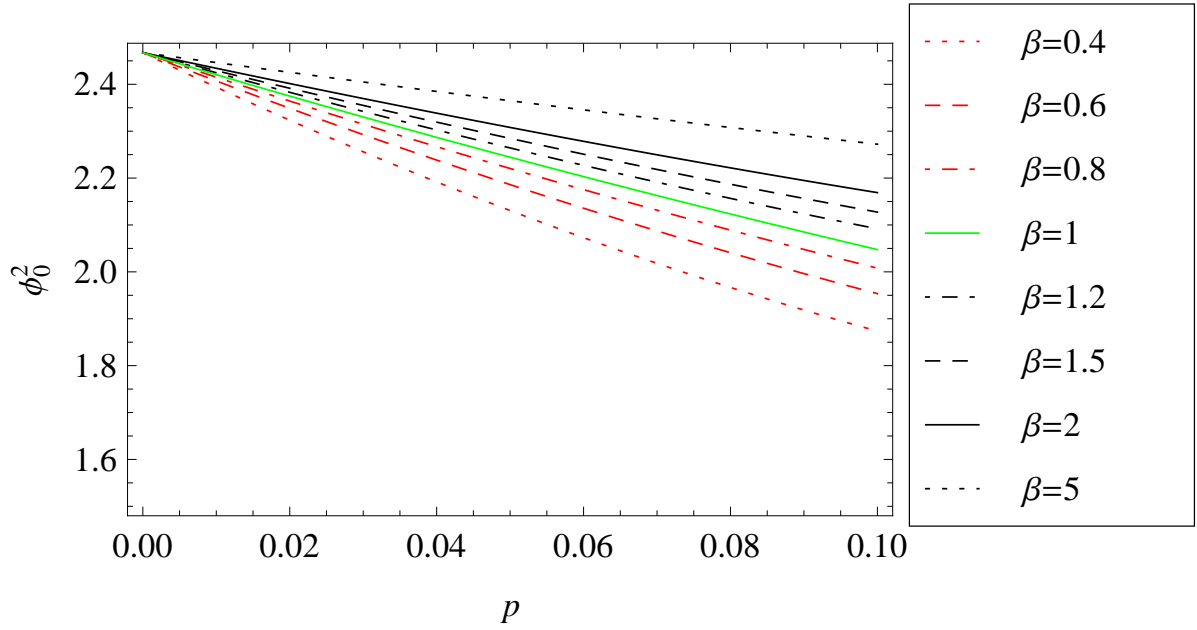}}
  \subfigure[]{\includegraphics[width=\textwidth]{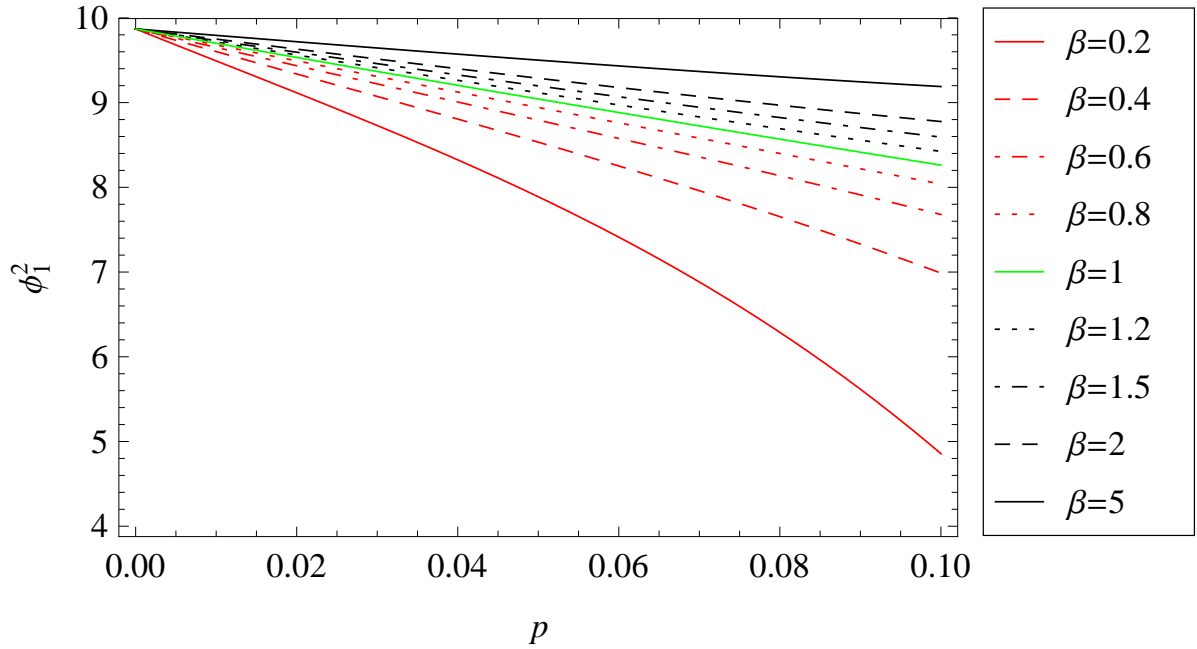}}
  \caption{The plot of $\Phi_0^2$ (a) and $\Phi_1^2$ (b), which are proportional to the energies $E_0$, $E_1$, as functions of the inverse potential strength $p$.}
\end{figure}

\begin{equation}
\tan \Phi\simeq\frac{0.45\Phi\left(  \Phi-n\pi\right)  }{2\Phi-\left(  2n-1\right)  \pi
},\ n\pi<\Phi<n\pi+\frac{\pi}{2}
\end{equation}

Even the "parabolic approximation" for $\left(  \cos \Phi/\Phi\right)  ^{2}$ or
$\left(  \sin \Phi/\Phi\right)  ^{2}$, in the sense used in Ref.$\ $ \cite{VB_RD}, gives a
quartic equation for the dimensionless wave vector. Even if a quartic equation
can be reduced to a cubic and a quadratic equation, the results are quite
complicated. So the only practically
acceptable approximations are the series expansions of the previous
"algebraizations" of $\cos\Phi,\ \tan\Phi,$ in terms of $\Phi/2n\pi
\sim1/4n,\ n\eqslantgtr2.$ So, the errors of a such approximation, which keeps only the first order term, will be about $(1/4n)^2$, $n\eqslantgtr2$.

\subsection{Graphical illustration of our main results}

In order to illustrate graphically some of our results, let us notice that,
using Eqs. \eqref{potential_strength_P}-\eqref{wave_vector_kout}, we can write the following relation for the energy:

\begin{equation}
\frac{m_{i}L^{2}}{2\hbar^{2}}E_{n}
=
\Phi_{n}^{2}=z_{n}\left(p\right)
\label{dimensionless_wave_vector_squared}
\end{equation}
where $z_{n}$ is the root of the cubic equations obtained after the
algebraization of the transcendental eigenvalue equations for the ground state$(n=0)$
and for the first excited state$(n=1)$. According to the Eqs. \eqref{sol_approx_even_smaller}, \eqref{sol_approx_odd_bigger_beta}, for a
deep well, the root $z$ (for simplicity, we dropped the index of $z$) can be
approximated with a quartic potential in $p$, the inverse of the potential
strength $P$. Let us mention that, if we replace in the definition of
$P$, Eq. \eqref{potential_strength_P}, $m_{i}$ with the free electron mass, we choose the length of the
well $L=10nm$ and we express the potential $V_{0}$ in electronvolts, we get:%

\begin{equation}
P=25.\,\allowbreak616\sqrt{V_{0}},\ p=3.\,\allowbreak903\,8\times10^{-2}%
\frac{1}{\sqrt{V_{0}}}
\end{equation}
Consequently, in this example, the well can be considered as "deep", and the
energy, given by \eqref{dimensionless_wave_vector_squared}, is actually a function of the depth of the well, $V_{0}$.

We shall plot our main results, i.e. the series expansions of the
dimensionless wave vectors, $\Phi_{0}^{2}$ and $\Phi_{1}^{2}$, as functions of
$p,$ on the range $0<p<0.1,$ when the conditions of convergence are
satisfactorily fulfilled. The energy is a monotonically increasing function of
$\beta;$ its values, for $\beta=1,$ are obtained from Eqs. \eqref{even_states}, \eqref{odd_states};
they are approximated with algebraic equations using the formulas \eqref{dABG_formula} and \eqref{sinc_approx}.

\section{Discussion: applications to other nanostructures}
Our calculations can be easily applied to type II heterointerfaces, when one of the effective mass of the charge carrier is negative: $m_{in}m_{out}<0$ (Ref. \cite{Bastard}, Ch. III, eqs. (35), (36)); a detailed description of such heterointerfaces can be found for instance in Ref. \cite{Bastard}, p.66. So, instead of \eqref{Bastard_even}, \eqref{Bastard_odd}, the eigenvalue equations take the form:
\begin{align}
  \tan \frac{k_{in,2n} L} {2}&=
        -\frac{m_i}{|m_o|} \frac{k_{out,2n}} {k_{in,2n}}
        =-|\beta|\frac{k_{out,2n}} {k_{in,2n}}, \mathrm{\ even \ states}\\
  \tan \frac{k_{in,2n+1}L} {2}&=
        \frac{|m_o|} {m_i} \frac{k_{in,2n+1}} {k_{out,2n+1}}
        =\frac{1} {|\beta|} \frac{k_{in,2n+1}} {k_{out,2n+1}}, \mathrm{\ odd \ states}
\end{align}
    and can be solved following exactly the same approach.

    As already mentioned, the wave function in the Schr\"{o}dinger equation \eqref{Schrodinger_eq} can be interpreted as an envelope function. This approximation works well when the materials constituting the heterostructures are perfectly lattice-matched and they crystallize in the same crystallographic structure (in most cases, the zinc blende structure). Its application is restricted to the vicinity of the high-symmetry points in the host's Brillouin zone ($\Gamma$, $X$, $L$). Actually, most of the heterostructures' energy levels relevant to actual devices are relatively close to a high symmetry point in the host's Brillouin zone. A popular example is given by the lowest conduction states of GaAs-GaAlAs heterostructures with GaAs layer thickness is about  100A or larger. A detailed description of the cases in which the envelope function model is successful is given in Ref. \cite{Bastard}, p. 68.

    As there are some similarities between quantum wells and quantum dots, our results are also relevant for these devices. The simplest remark is that the eigenvalues equation for the first odd state in a quantum well is identical to that corresponding to the $l=0$ state in a quantum dot (see for instance Ref. \cite{Fluegge}, pr. 63). Also, the eigenvalue equations for the wave vectors of bound energy levels of a finite barrier rectangular shaped quantum dot, Eq. (36) in Ref. \cite{Ata}, are quite similar to ours - \eqref{even_states_tan}, \eqref{odd_states_cot}, but somewhat more complicated. The ground state energy of electrons and holes in a core/shell quantum dot is given by Eq. (21) of Ref. \cite{Ibral}, an equation similar to ours, just mentioned previously. Such results are important, inter alia, for the interpretation of photoluminescence spectra and photon harvesting of quantum dots.

\section{Conclusions}
     In this paper, we solved a simple problem: the Schr\"{o}dinger equation
for a square potential, with BenDaniel - Duke boundary conditions,
paying special attention to approximate analytical expressions of
energy eigenvalues. All the results are obtained with mathematics
accessible to an undergraduate student. This simple exercise can
facilitate student's understanding of more complex and profound
issues, like position-dependent mass, envelope function or type I and
II semiconductor heterojunctions. Also, it valorizes the simplest way
of solving the cubic equation, a subject usually avoided at
undergraduate level.
    Our approach can be extended to quantum dots of various architectures
and can be used for explanation of photoabsorbtion spectra and photon
harvesting.

\bigskip

\appendix

\section{The real roots of a cubic equation}

We shall outline here some basic aspects concerning the cubic equation \cite{Wolfram}. Although it was solved more than five centuries ago - in circa 1500 by Scipione del Ferro, the most convenient form of the solutions, obtained using hypergeometric functions, became popular quite late \cite{Zucker}. Our intention is to give here a short and practical guide for obtaining the real roots of a cubic equation.

 Let us
consider its general form:%
\begin{equation}
z^{3}+a_{2}z^{2}+a_{1}z+a_{0}=0
\label{cubic_eq_z}
\end{equation}
With a change of variable:%
\begin{equation}
z=x-\frac{1}{3}a_{2}
\end{equation}
it can be put in the "depressed form":%
\begin{equation}
x^{3}+p_{c}x=q_{c}
\label{depressed_eq}
\end{equation}
where:%
\begin{equation}
p_{c}=\frac{3a_{1}-a_{2}^{2}}{3},\ q_{c}=\frac{9a_{1}a_{2}-27a_{0}-2a_{2}^{3}%
}{27}
\end{equation}

The standard notations for these quantities is $p,\ q.$ In order to avoid the
confusion with the inverse of the potential strength, $p,$ we used the symbol
$p_{c}$ (the index $c$ is for "cubic"). With a scale transformation:%
\begin{equation}
x=\sqrt{\frac{4\left\vert p_{c}\right\vert }{3}}y
\end{equation}
the depressed equation \eqref{depressed_eq} can be written in a form appropriate for
comparison with triple-angle formulae for trigonometric or hyperbolic functions:%
\begin{equation}
4y^{3}+3\sigma_{p}y=K
\end{equation}
where%
\begin{equation}
K=\frac{1}{2}q_{c}\left(  \frac{3}{\left\vert p_{c}\right\vert }\right)
^{3/2}
\end{equation}
and $\sigma_{p}$ is the signum function:%
\begin{equation}
\sigma_{p<0}=-1,\ \sigma_{p>0}=+1
\end{equation}

If $p_{c}>0,$ the equation:%
\begin{equation}
4y^{3}+3y=K
\label{cubic_y_eq}
\end{equation}
has one real root:%
\begin{equation}
y=\sinh\left(  \frac{1}{3}\sinh^{-1}K\right)
\label{sol_unique_y}
\end{equation}

If $p_{c}<0$ and $\left\vert K\right\vert \geq1$, the equation%
\begin{equation}
4y^{3}-3y=K,\ \left\vert K\right\vert \geq1
\end{equation}
has only one real root:%
\begin{equation}
y\left(  K\geq1\right)  =\cosh\left(  \frac{1}{3}\cosh^{-1}K\right)
\label{sol_unique_Kbigger1}
\end{equation}
or%
\begin{equation}
y\left(  K\leq-1\right)  =-\cosh\left(  \frac{1}{3}\cosh^{-1}\left\vert
K\right\vert \right)
\label{sol_unique_Ksmaller-1}
\end{equation}

If $p_{c}<0$ and $\left\vert K\right\vert \leq1$, the equation%
\begin{equation}
4y^{3}-3y=K,\ \left\vert K\right\vert \leq1
\end{equation}
has three real roots:%
\begin{equation}
y_{1}=\sin\left(  \frac{1}{3}\arcsin K-\frac{\pi}{3}\right)
\label{sol_three_first}
\end{equation}

\begin{equation}
y_{2}=\sin\left(  \frac{1}{3}\arcsin K+\frac{\pi}{3}\right)
\label{sol_three_second}
\end{equation}

\begin{equation}
y_{3}=-\sin\left(  \frac{1}{3}\arcsin K\right)
\label{sol_three_third}
\end{equation}
Finally, the roots of the general equation \eqref{cubic_eq_z} are:%
\begin{equation}
z=x-\frac{1}{3}a_{2}=\sqrt{\frac{4\left\vert p_{c}\right\vert }{3}}y-\frac
{1}{3}a_{2}
\label{sol_cubic_eq}
\end{equation}

So, the algorithm to be followed in order to solve the cubic equation is the following:

1. we calculate $p_{c},\ q_{c}$ and determine their sign;

2a. if $p_{c}>0,$ the equation \eqref{cubic_y_eq} has only one real root, Eq. \eqref{sol_unique_y};

2b. if $p_{c}<0,$ we check if $\left\vert K\right\vert $ is $\lessgtr1;$

2b.1 if $\left\vert K\right\vert >1,$ the equation \eqref{cubic_eq_z} has only one real
root, given by Eq. \eqref{sol_unique_Kbigger1} or Eq. \eqref{sol_unique_Ksmaller-1}, according to the sign of $K.$

2b.2 if $\left\vert K\right\vert <1,$ the equation \eqref{cubic_eq_z} has three real roots,
given by Eqs. \eqref{sol_three_first},\eqref{sol_three_second} and \eqref{sol_three_third}.

\bigskip

Using elementary trigonometric identities, the equations \eqref{sol_three_first}, \eqref{sol_three_second} and \eqref{sol_three_third} can be put
in the more elegant form:%
\begin{equation}
t_{k}=\cos\left(  \frac{1}{3}\arccos K-\frac{2\pi k}{3}\right)
\end{equation}
It is easy to check that:%
\begin{equation}
t_{0}=y_{2},\ t_{1}=y_{3},\ t_{2}=y_{1}
\end{equation}

The case of multiple roots will be discussed separately. If we make Viete's substitution:%
\begin{equation}
x=w-\frac{p_{c}}{3w}%
\label{Vietes_substitution}
\end{equation}
in the depressed equation, we get a quadratic equation in $w^{3}:$%
\begin{equation}
\left(  w^{3}\right)^{2}-q_{c}w^{3}-\frac{1}{27}p_{c}^{3}%
=0%
\end{equation}
with the roots:%
\begin{equation}
w^{3}=\frac{1}{2}q_{c}\pm\sqrt{\frac{1}{4}q_{c}^{2}+\frac{1}{27}p_{c}^{3}%
}=R\pm\sqrt{R^{2}+Q^{3}}%
\end{equation}

The existence of a multiple root corresponds to the condition:%
\begin{equation}
\frac{1}{4}q_{c}^{2}=-\frac{1}{27}p_{c}^{3}
\end{equation}
this means that:%
\begin{equation}
p_{c}<0,\ K^{2}=1
\end{equation}
The value of the multiple (double) root is obtained putting $w^{3}=\left(
q_{c}/2\right)  ^{1/3}$ in Viete's substitution Eq. \eqref{Vietes_substitution}.

\bigskip

\begin{acknowledgments}
The authors acknowledge the financial support of the IFIN-HH - ANCSI project PN 16 42 01 01/2016.
\end{acknowledgments}

\end{document}